\documentclass[letterpaper]{elsart5p}
\advance\voffset-45pt
\advance\hoffset8.075pt

\usepackage[dvips]{pstcol}
\usepackage{graphicx}
\usepackage{dcolumn}
\usepackage{bm}
\usepackage{rotating}
\usepackage{amsmath,amssymb,amsfonts}
\usepackage{pifont}
\usepackage[normalem]{ulem}
\usepackage[normal]{subfigure}

\newcommand{\hermes}{{\sc Hermes}} 
\newcommand{\compass}{{\sc Compass}}
\newcommand{\belle}{{\sc Belle}}

\newcommand{\desy}{{\sc Desy}}
\newcommand{\hera}{{\sc Hera}}

\newcommand{\jetset}{{\sc Jetset}}
\newcommand{\pythia}{{\sc Pythia6}}

\newcommand{\cmh}{\mbox{\large$\bm{\langle}$}\ensuremath{\sin(\phi\!+\!\phi_S)}
        \mbox{\large$\bm{\rangle}$}^h_{{\mathrm U}{\mathrm T}}}

\newcommand{\pperp}{P_{h\perp}}

\begin{document}

\date{ \today}

\begin{frontmatter}

\title{Effects of transversity in deep-inelastic scattering by polarized protons}

\collab{\hermes\ Collaboration}

\journal{Physics Letters B}

\author[12,15]{A.~Airapetian},
\author[26]{N.~Akopov},
\author[5]{Z.~Akopov},
\author[6]{E.C.~Aschenauer\thanksref{27}},
\author[25]{W.~Augustyniak},
\author[26]{R.~Avakian},
\author[26]{A.~Avetissian},
\author[5]{E.~Avetisyan},
\author{A.~Bacchetta\thanksref{28}},
\author[18]{S.~Belostotski},

% ...
\address[5]{DESY, 22603 Hamburg, Germany}
\address[6]{DESY, 15738 Zeuthen, Germany}
% ...
\address[12]{II. Physikalisches Institut, Universit\"at Gie{\ss}en, 35392 Gie{\ss}en, Germany}
% ...
\address[15]{Randall Laboratory of Physics, University of Michigan, Ann Arbor, Michigan 48109-1040, USA }
% ...
\address[18]{Petersburg Nuclear Physics Institute, Gatchina, Leningrad region, 188300 Russia}
% ...
\address[25]{Andrzej Soltan Institute for Nuclear Studies, 00-689 Warsaw, Poland}
\address[26]{Yerevan Physics Institute, 375036 Yerevan, Armenia}

\thanks[27]{Now at: Brookhaven National Laboratory, Upton, NY 11772-5000, USA}
\thanks[28]{Address: Dipartimento di Fisica Nucleare e Teorica, Universit\`a di Pavia and Istituto Nazionale di Fisica Nucleare, Sezione di Pavia, via Bassi 6, 27100 Pavia, Italy}
\thanks[29]{Now at: Massachusetts Institute of Technology, Cambridge, MA 02139, USA}
\thanks[30]{Now at: Institut f\"ur Physik, Universit\"at Mainz, 55128 Mainz, Germany}
\thanks[31]{Now at: Institut f\"ur Kernphysik, Universit\"at Frankfurt a.M., 60438 Frankfurt a.M., Germany}
\thanks[32]{Now at: Department of Physics, Yamagata University, Kojirakawa-cho 1-4-12, Yamagata 990-8560, Japan}\thanks[33]{Now at: Carnegie Mellon University, Pittsburgh, PA 15213, USA}

\begin{abstract}
Single-spin asymmetries for pions and charged kaons are measured in semi-inclusive deep-inelastic scattering of positrons and electrons off a transversely nuclear-polarized hydrogen target. The dependence of the cross section on the azimuthal angles of the target polarization $(\phi_S)$ and the produced hadron $(\phi)$ is found to have a substantial \(\sin(\phi+\phi_S)\) modulation for the production of $\pi^+$, $\pi^-$ and $K^+$. This Fourier component can be interpreted in terms of non-zero transversity distribution functions and non-zero favored and disfavored Collins fragmentation functions with opposite sign. For  \(\pi^0\) and \( K^- \) production the amplitude of this Fourier component is consistent with zero.
\end{abstract}


\end{frontmatter}

\maketitle


Most of our knowledge about the internal structure of nucleons comes from deep-inelastic scattering (DIS) experiments. At the energies of current fixed-target experiments, the dominant process in DIS of charged leptons by nucleons is the exchange of a single space-like photon with a squared four-momentum $-Q^2$ much larger than the typical hadronic scale, usually set to be the squared mass $M^2$ of the nucleon. The cross section for this lepton scattering process can be decomposed in a model-independent way in terms of structure functions. Factorization theorems based on quantum chromodynamics (QCD) provide an interpretation of these structure functions in terms of parton distribution functions (PDFs), which ultimately reveal crucial aspects of the dynamics of confined quarks and gluons.

\begin{figure}[t]
\centering
\includegraphics[height=3.7cm,angle=0]{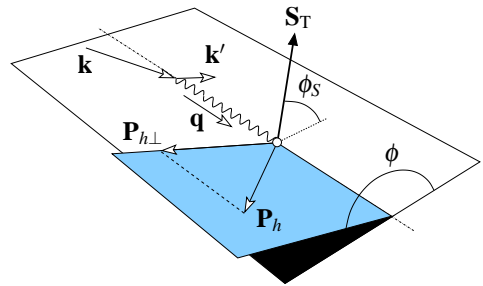}
\caption{\label{fig:angles} The definition of the azimuthal angles $\phi$ and $\phi_S$ relative to the lepton scattering plane.}
\end{figure}

Polarized inclusive DIS on nucleons, $lN\rightarrow l'X$ (where $X$ denotes the undetected final state), neglecting weak boson exchange can be described by four structure functions (see, e.g, Refs.~\cite{Jaffe:1989xx,Lampe:1998eu}). They can be interpreted using collinear factorization theorems (see, e.g, Ref.~\cite{Collins:1989gx,Brock:1993sz} and references therein). Three of the structure functions contain contributions at leading order in an expansion in $M/Q$ (twist expansion). These contributions include the leading-twist (twist-2) quark distribution functions $f_1^q(x)$ and $g_1^q(x)$~\cite{Lampe:1998eu} (for simplicity, the dependence on $Q^2$ has been dropped). The variable $x$ represents the fraction of the nucleon momentum carried by the parton in a frame where the nucleon moves infinitely fast in the direction opposite to the probe. The hard probe defines a specific direction ($\bf{q}$ in Fig.~\ref{fig:angles}), usually denoted as longitudinal, and the transverse plane perpendicular to it. In a parton-model picture, $f_1^q(x)$ describes the number density of quarks of flavor $q$ in a fast-moving nucleon without regard to their polarization. The PDF $g_1^q(x)$ describes the difference between the number densities of quarks with helicity equal or opposite to that of the nucleon if the nucleon is longitudinally polarized. The integrals over $x$ of $f_1^q(x)$ and $g_1^q(x)$ are related to the vector and axial charge of the nucleon, respectively.

There is a third leading-twist PDF, the function $h_1^q(x)$\footnote{In literature, the distribution functions $f_1^q(x)$, $g_1^q(x)$, and $h_1^q(x)$ are also denoted as $q(x)$, $\Delta q(x)$, and $\delta q(x)$, respectively.}, 
called the transversity distribution (see Ref.~\cite{Barone:2001sp} for a review on the subject). Its integral over $x$ is related to the tensor charge of the nucleon~\cite{He:1994gz}. It can be interpreted as the difference between the densities of quarks with transverse (Pauli-Lubanski) polarization parallel or anti-parallel to the transverse polarization of the nucleon~\cite{Jaffe:1991kp}. In contrast to $f_1^q(x)$ and $g_1^q(x)$, due to helicity conservation, there exist no gluon analog of $h_1^q(x)$ in the case of spin-$\frac{1}{2}$ targets. Therefore, $h_1^q(x)$ cannot mix with gluons under QCD evolution.

The transversity distribution does not appear in any structure function in inclusive DIS because it is odd under inversion of the quark chirality. It must be combined with another chiral-odd nonperturbative partner to appear in a cross section for hard processes involving only QED or QCD, as such interactions preserve chirality. For this reason, in spite of decades of inclusive DIS studies, no experimental information on the transversity distribution was available until recently.
In lepton-nucleon scattering, the transversity distribution can be accessed experimentally only in semi-inclusive DIS with a transversely polarized target, where it can appear in combination with, e.g., the chiral-odd  Collins fragmentation function~\cite{Collins:1992kk}. This Letter presents a measurement of the associated signal. 

In semi-inclusive DIS, $lN\rightarrow l'hX$, where a hadron $h$ is detected in the final state in coincidence with the scattered lepton, the cross section depends on, among other variables, the hadron transverse momentum and its azimuthal orientation with respect to the lepton scattering plane about the virtual-photon direction. If the target is polarized and the polarization of the final state is not measured, the semi-inclusive DIS cross section can be decomposed in terms of 18 semi-inclusive structure functions (see, e.g, Ref.~\cite{Bacchetta:2006tn}).

When the transverse momentum of the produced hadron is small compared to the hard scale $Q$, semi-inclusive DIS can be described using transverse-momentum-dependent factorization~\cite{Collins:1981uk,Ji:2004wu}. The semi-inclusive structure functions can be interpreted in terms of convolutions involving transverse-momentum-dependent parton distribution and fragmentation functions~\cite{Collins:1982uw}. The former encode information about the distribution of partons in a three-dimensional momentum space, and the latter describe the hadronization process in a three-dimensional momentum space. Hence, the study of semi-inclusive DIS not only opens the way to the measurement of transversity, but also probes new dimensions of the structure of the nucleon and of the hadronization process, thus offering new perspectives to our understanding of QCD.

When performing a twist expansion, eight semi-inclusive structure functions contain contributions at leading order, related to the eight leading-twist transverse-momentum-dependent PDFs~\cite{Bacchetta:2006tn}. One of these structure functions is interpreted as the convolution of the transversity distribution function $h_1^q(x,p_{\mathrm T}^2)$ (not integrated over the transverse momentum) and the Collins fragmentation function $H_{1}^{\perp q\rightarrow h}(z,k_{\mathrm T}^2)$, which acts as a polarimeter being sensitive to the correlation between the transverse polarization of the fragmenting quark and $\bf{k}_{\mathrm T}$ \cite{Collins:1992kk}. Here, $z$ in the target-rest frame denotes the fraction of the virtual photon energy carried by the produced hadron $h$, \({\bf p}_{\mathrm T}\) denotes the transverse momentum of the quark with respect to the parent nucleon direction, and \({\bf k}_{\mathrm T}\) denotes the transverse momentum of the fragmenting quark with respect to the direction of the produced hadron. This structure function manifests itself as a $\sin(\phi+\phi_S)$ modulation in the semi-inclusive DIS cross section with a transversely polarized target. Its Fourier amplitude, henceforth named Collins amplitude, is denoted as \(2\cmh\), where $\phi$ ($\phi_S$) represents the azimuthal angle of the hadron momentum (of the transverse component of the target spin) with respect to the lepton scattering plane and about the virtual-photon direction, in accordance with the {\it Trento Conventions}~\cite{Bacchetta:2004jz} (see Fig.~\ref{fig:angles}). The subscript UT denotes unpolarized beam and target polarization transverse with respect to the virtual-photon direction. Other azimuthal modulations have different origins and involve other distribution and fragmentation functions. They can be disentangled through their specific dependence on the two azimuthal angles $\phi$ and $\phi_S$ (see, e.g, Refs.~\cite{Bacchetta:2006tn,Mulders:1996dh,Boer:1998nt}). Results on, e.g., the  $\sin(\phi-\phi_S)$ modulation of this data set were reported in Ref.~\cite{Airapetian:2009ti}.

Non-zero Collins amplitudes were previously published for charged pions from a hydrogen target~\cite{Airapetian:2004tw}, based on a small subset (about $10\%$) of the data reported here, consisting of about 8.76 million DIS events. Collins amplitudes 
for unidentified hadrons were measured on protons~\cite{Alekseev:2010rw} 
and for pions and kaons, albeit consistent with zero, 
on deuterons~\cite{Alexakhin:2005iw,Ageev:2006da,Alekseev:2008dn} by the \compass\ collaboration. 
In Refs.~\cite{Anselmino:2007fs,Anselmino:2008jk} the first joint extraction of the transversity distribution function and the Collins fragmentation function was carried out, under simplifying assumptions, using preliminary results from a subset of the present data in combination with the deuteron data from the \compass\ collaboration~\cite{Alexakhin:2005iw,Ageev:2006da,Alekseev:2008dn} and $e^+e^-$ annihilation data from  the \belle\ collaboration~\cite{Abe:2005zx,Seidl:2008xc}.
Recently, significant amplitudes for two-hadron production in semi-inclusive DIS, which constitutes an independent process to probe transversity, were measured at the \hermes\ experiment \cite{Airapetian:2008sk} providing additional evidence for a non-zero transversity distribution function.

In this Letter, in addition to much improved statistical precision on the charged pion results, the Collins amplitudes for identified $K^+$, $K^-$, and $\pi^0$ are presented for the first time for a proton target.
The data reported here were recorded during the 2002--2005 running period of the \hermes\ experiment with a transversely nuclear-polarized hydrogen target stored in an open-ended target cell internal to the \(27.6\)\,GeV  \hera\ polarized positron/electron storage ring at \desy. The two beam helicity states are almost perfectly balanced in the present data, and no measurable contribution arising from the residual net beam polarization to the amplitudes extracted was observed. The target cell was fed by an atomic-beam source~\cite{Nass:2003mk}, which uses Stern--Gerlach separation combined with radio-frequency transitions of hyperfine states. The target cell was immersed in a transversely oriented magnetic holding field. The effects of this magnetic field were taken into account in the reconstruction of the vertex positions and the scattering angles of charged particles. The nuclear polarization of the atoms was flipped at 1--3\, minutes time intervals, while both the polarization and the atomic fraction inside the target cell were continuously measured~\cite{Airapetian:2004yf}. The average magnitude of the proton-polarization component perpendicular to the beam direction was $0.725\pm 0.053$.
Scattered leptons and coincident hadrons were detected by the \hermes\ spectrometer~\cite{Ackerstaff:1998av}. 
Leptons were identified with an efficiency exceeding 98\% and a hadron contamination of less than 1\%. Charged hadrons detected within the momentum range 2--15 GeV were identified using a dual-radiator RICH by means of a hadron-identification algorithm that takes into account the event topology. The detection of the neutral pions is based on the measurements of photon pairs in the electromagnetic calorimeter. These were accepted only if $E_{\gamma}>1$ GeV and $0.10~{\rm GeV} < M_{\gamma \gamma} < 0.17~{\rm GeV}$, where $E_{\gamma}$ and $M_{\gamma \gamma}$ denote the photon energy and the photon-pair invariant mass, respectively. The combinatorial background was evaluated in the side-bands $0.06~{\rm GeV} < M_{\gamma \gamma} < 0.10~{\rm GeV}$ and $0.17~{\rm GeV} < M_{\gamma \gamma} < 0.21~{\rm GeV}$.

Events were selected according to the kinematic requirements $W^2 >10$\,GeV$^2$, $0.023<x< 0.4$, $0.1 < y <  0.95$, and $Q^2>1$\,GeV$^2$, where $W^2\equiv(P+q)^2$, \(Q^2\equiv -q^2\equiv -(k-k')^2\), \(y\equiv (P\cdot q)/(P\cdot k)\), and  \(x\equiv Q^2/(2P\cdot q)\) are the conventional DIS kinematic variables with \(P\), $k$ and $k'$ representing the four-momenta of the initial state target proton, incident and outgoing lepton, respectively. In order to minimize target fragmentation effects as well as to exclude kinematic regions where contributions from exclusive channels become sizable, coincident hadrons were only included if \(0.2<z<0.7\), where \(z\equiv(P\cdot P_h)/(P\cdot q)\) and $P_h$ is the four-momentum of the produced hadron.

The cross section for semi-inclusive production of hadrons using an unpolarized lepton beam and a target polarized transversely with respect to the virtual photon direction includes a polarization-averaged part and a polarization-dependent part. The former contains two cosine modulations and the latter contains a total of five sine modulations \cite{Bacchetta:2006tn,Mulders:1996dh,Boer:1998nt}: 
\begin{equation} 
\begin{split} 
d\sigma^h(\phi,\phi_S) &= d\sigma^h_{\mathrm{UU}} \biggr\{1+\sum_{n=1}^2 2 \langle \cos (n \phi)
\rangle^h_{\mathrm{UU}} \cos (n \phi)
\\
&\quad 
+ |{\bf S}_{\mathrm T}| \sum_{i=1}^5 2 \langle \sin \Phi_i
\rangle^h_{\mathrm{UT}} \sin \Phi_i \biggl\},
\label{eq:sigma_UT}
\end{split} 
\end{equation} 
where
${\bf S}_{\mathrm T}$ denotes the transverse (with respect to the virtual photon direction) component of the target-proton polarization vector and \(\Phi=[\phi+\phi_S,\phi-\phi_S,\phi_S,2 \phi-\phi_S,3 \phi-\phi_S]\). The dependence of the cross section and of the azimuthal amplitudes on $x$, $y$, $z$, and $P_{h\perp}$ has been suppressed. The subscript UU denotes unpolarized beam and unpolarized target, and \(d\sigma^h_{\mathrm{UU}}\) represents the cross section averaged over $\phi$ and over beam and target polarizations. 

The Collins amplitude \(2\cmh\) can be interpreted in the parton model as \cite{Mulders:1996dh}
\begin{equation}
\begin{split} 
&2\cmh (x,y,z,P_{h \perp})
\\ 
&=\frac{(1-y)}{(1-y+y^2/2)}\frac{{\cal C}\bigl[ - \frac{{\bf P}_{h\perp} \cdot {\bf k}_{\mathrm T}}{|{\bf P}_{h\perp}|~ M_h} \,h_1^q(x,p_{\mathrm T}^2) H_1^{\perp q \rightarrow h}(z,k_{\mathrm T}^2)\bigr]}{{\cal C}\bigl[ f_1^q(x,p_{\mathrm T}^2) D_1^{q \rightarrow h}(z,k_{\mathrm T}^2)\bigr]},
\label{eq:QPM-collins}
\end{split} 
\end{equation}
where \({\bf P}_{h\perp}\equiv | {\bf P}_h - \frac{({\bf P}_h \cdot {\bf q}){\bf q}}{|{\bf q}|^2}| \) is the transverse momentum of the produced hadron, and \(D_1^{q\rightarrow h}\) is the polarization-averaged quark fragmentation function. The notation ${\cal C}$ denotes the convolution~\cite{Bacchetta:2006tn} 
\begin{equation} 
{\cal C} \bigl[ {\it ...} \bigr] =
x {\sum_q} e_q^2 \int d^2 {\bf p}_{\mathrm T} d^2 {\bf k}_{\mathrm T} \, \delta^{(2)} \biggl({\bf p}_{\mathrm T} - {\bf k}_{\mathrm T} -\frac{{\bf P}_{h\perp}}{z} \biggr) \bigl[{\it ...}\bigr] ~,
\label{eq:convolution}
\end{equation} 
where the sum runs over the quark flavors $q$, and \(e_q\) are the quark electric charges in units of the elementary charge. Expressions similar to Eq.~(\ref{eq:QPM-collins}) hold for the other azimuthal modulations in Eq.~(\ref{eq:sigma_UT}) \cite{Bacchetta:2006tn}. Note that, as the quark flavors enter the cross section with the square of their electric charge, the $u$-quarks provide the dominant contribution to the production of, e.g., $\pi^+/K^+$ for proton targets (commonly denoted as ``$u$-quark dominance'').

Experimentally, the Fourier amplitudes of the yields for opposite transverse target-spin states were extracted using a maximum-likelihood fit alternately binned in $x$, $z$, and $P_{h\perp}$, but unbinned in $\phi$ and $\phi_S$. This is equivalent to a Fourier decomposition of the asymmetry
\begin{equation} 
A_{\mathrm{UT}}^h(\phi,\phi_S) \equiv \frac{1}{|{\bf S}_{\mathrm T}|} \frac{d\sigma^h(\phi,\phi_S)-d\sigma^h(\phi,\phi_S+\pi)}{d\sigma^h(\phi,\phi_S)+d\sigma^h(\phi,\phi_S+\pi)}~,
\label{eq:asymmetry}
\end{equation}  
for perfectly balanced target polarization and in the limit of very small $\phi$ and $\phi_S$ bins. The asymmetry amplitudes for neutral pions were corrected for the effects of the combinatorial background evaluated in the side-bands of the photon-pair invariant mass spectrum. In addition to the five sine terms in Eq.~(\ref{eq:sigma_UT}), the fit also included a ${\rm sin(2\phi+\phi_S)}$ term, arising from the small but non-vanishing target-polarization component that is longitudinal to the virtual-photon direction when the target is polarized perpendicular to the beam direction~\cite{Diehl:2005pc}. In order to avoid cross contamination arising from the limited spectrometer acceptance, the six amplitudes were extracted simultaneously. The fit did not include the ${\rm cos}(n\phi)$ modulations of Eq.~\eqref{eq:sigma_UT}. As a consequence, one cannot expect {\it a priori} that the Fourier amplitudes extracted are identical to those of Eq.~(\ref{eq:sigma_UT}). However, in the following they will be considered to be equivalent because inclusion in the fit of estimates \cite{Giordano:2009hi} for the \(\cos(\phi)\) and \(\cos(2\phi)\) amplitudes of the unpolarized cross section resulted in negligible effects on the extracted amplitudes.

\begin{figure}
\centering
\includegraphics[width=0.485\textwidth,angle=0]{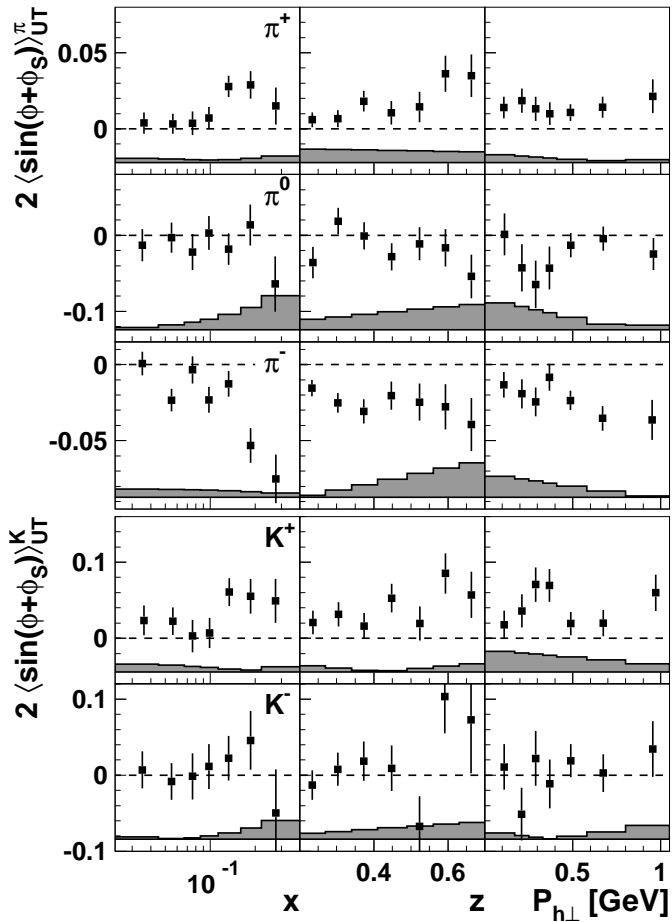}
\caption{\label{fig:main-results-all} Collins amplitudes for pions and charged kaons as a function of \(x\), \(z\), or \(\pperp\). The systematic uncertainty is given as a band at the bottom of each panel. In addition there is a 7.3\% scale uncertainty from the accuracy in the measurement of the target polarization.}
\end{figure}

The extracted Collins amplitudes are shown in Fig.~\ref{fig:main-results-all} as a function of \(x\), \(z\), or \(P_{h\perp}\). They are positive for \(\pi^+\) and \(K^+\), negative for \(\pi^-\), and consistent with zero for \(\pi^0\) and \(K^-\) at a confidence level of at least $95\%$ based on a Student's {\it t}-test including the systematic uncertainties. Note that the $x$, $z$, and $P_{h\perp}$ dependences in Fig.~\ref{fig:main-results-all} are three projections of the same data and are thus fully correlated.

A scale uncertainty of 7.3\% on the extracted amplitudes, not shown in Fig.~\ref{fig:main-results-all}, arises from the accuracy in the measurement of the target polarization. Effects from acceptance, smearing due to detector resolution, higher order QED processes and hadron identification procedure based on the RICH are not corrected for in the data. Rather, the size of all these effects was estimated using a \pythia\ Monte Carlo simulation~\cite{Sjostrand:2000wi} tuned to \hermes\ hadron multiplicity data and exclusive vector-meson production data~\cite{Liebing:2004us,Hillenbrand:2005ke,Maiheu:thesis} and including a full simulation of the \hermes\ spectrometer. A polarization state was assigned to each generated event using a model that reflects the (transverse target) polarization dependent part of the cross section (see Eq.~(\ref{eq:sigma_UT})). This model was obtained through a fully differential (i.e differential in the four relevant kinematic variables $x$, $Q^2$, $z$, and $P_{h\perp}$) $2^{nd}$ order polynomial fit~\cite{Pappalardo:2008zza,Diefenthaler:thesis} of real data. The asymmetry amplitudes, extracted from the simulated data by means of the same analysis procedure used for the real data, were then compared with the model, evaluated in each bin at the mean kinematics, to obtain an estimate of the global impact of the effects listed above. The result was included in the systematic uncertainty and constitutes the largest contribution. It accounts for effects of nonlinearity of the model, as it includes the difference in each bin between the average model and the model evaluated at the average kinematics.  
The impact on the extracted amplitudes of contributions \cite{Diehl:2005pc} from the non-vanishing longitudinal target-spin component was estimated based on previous measurements of single-spin asymmetries for longitudinally polarized protons \cite{Airapetian:1999tv,Airapetian:2001eg}. The resulting relatively small effect was included in the systematic uncertainty.


A Monte Carlo simulation was used to estimate the fraction of pions and kaons originating from the decay of exclusively produced vector mesons, updating previous results reported in Ref.~\cite{Elschenbroich:2006vp}. For charged pions, this fraction is dominated by the decay of \(\rho^0\) mesons and, in the kinematic region covered by the present analysis, is of the order of 6-7\%. The vector-meson fractions for neutral pions and charged kaons are of the order of 2-3\%. The \(z\) and \(\pperp\) dependences of the fraction of pions and kaons stemming from the decay of exclusively produced vector mesons are shown in~\cite{Airapetian:2009ti} for the two kinematic regions $Q^2 < 4~\mathrm{GeV}^2$ and $Q^2 > 4~\mathrm{GeV}^2$ (the $x$ dependence was not reported due to the strong correlation between $x$ and $Q^2$ in the data). They exhibit maxima at high $z$ and low \(\pperp\). These contributions are considered part of the signal and were not used to correct the pion and kaon yields analysed in the present work. However, this information can be useful for the interpretation of the results.

In general, the non-vanishing amplitudes shown in Fig.~\ref{fig:main-results-all} increase in magnitude with $x$. This is consistent with the expectation that transversity mainly receives contributions from the valence quarks. A non negligible contribution from the sea quarks cannot be excluded, but is not expected to be large due to the fact that transversity cannot be generated in gluon splitting. The amplitudes are also found to increase with $z$, in qualitative agreement with the results for the Collins fragmentation function from the \belle\ experiment~\cite{Abe:2005zx,Seidl:2008xc}.
The results of Fig.~\ref{fig:main-results-all} {also} show that the \(\pi^-\) amplitude is of opposite sign to that of \(\pi^+\) and larger in magnitude. A possible explanation is dominance of \(u\) flavor among struck quarks, in conjunction with a substantial magnitude with opposite sign of the disfavored Collins fragmentation function describing, e.g, the fragmentation of $u$ quarks into $\pi^-$ mesons, as already suggested in Ref.~\cite{Airapetian:2004tw}. Opposite signs for the favored and disfavored Collins fragmentation functions are not in contradiction to the \belle\ results~\cite{Abe:2005zx,Seidl:2008xc} and are supported by the combined fits reported in~\cite{Anselmino:2007fs}. They can be understood in light of the string model of fragmentation~\cite{Artru:1995bh} (and also of the Sch\"afer--Teryaev sum rule~\cite{Schafer:1999kn}). If a favored pion is created at the string end by the first break, a disfavored pion from the next break is likely to inherit transverse momentum in the opposite direction. The string fragmentation model, the base of the successful and widespread \jetset\ generator~\cite{Ingelman:1996mq}, predicts such a $P_{h\perp}$ strong negative correlation between favored and disfavored pions.

Under the assumption of isospin symmetry, the fragmentation functions for neutral pions are assumed equal to the average of those for charged pions. Factorization of the semi-inclusive cross section results in the following isospin relation for the Collins amplitudes for pions:
\begin{equation} 
\begin{split} 
\langle \sin (\phi+\phi_S) \rangle_{\mathrm{UT}}^{\pi^+}+C\langle \sin (\phi+\phi_S) \rangle_{\mathrm{UT}}^{\pi^-}
\\
&\quad 
\hspace{-3.5cm}
-(1+C)\langle \sin (\phi+\phi_S) \rangle_{\mathrm{UT}}^{\pi^0}=0~,
\label{eq:isospin}
\end{split} 
\end{equation} 
where $C$ denotes the ratio of the polarization-averaged cross sections for semi-inclusive charged-pion production ($C\equiv d\sigma_{\mathrm{UU}}^{\pi^-}/d\sigma_{\mathrm{UU}}^{\pi^+}$). The extracted pion amplitudes are consistent with Eq.~(\ref{eq:isospin}).

The Fourier amplitudes for \(K^+\) are found to be larger than those for \(\pi^+\) at a confidence level of at least $90\%$ ($99\%$) based on a Student's {\it t}-test including (not including) the systematic uncertainties. On the other hand, the amplitudes for \(\pi^-\) and \(K^-\) exhibit a very different behavior, the former being significantly negative, while the latter is consistent with zero in the whole kinematic range. Here, however, one should keep in mind that, in contrast to $\pi^-$, a $K^-$ has no valence quarks in common with the target proton and sea quark transversity is expected to be small.

In interpreting the various features of the extracted amplitudes, and in particular the differences between those of pions and kaons, the largely unknown role of several concurring factors should be considered. Among these are, e.g, (i) the role of sea quarks in conjunction with possibly large fragmentation functions; (ii) the various contributions from decay of semi-inclusively produced vector-mesons which, based on a Monte Carlo simulation, are mainly $\rho$ and $\omega$ mesons producing pions (up to $37\%$ and $10\%$, respectively), and $K^*$ and $\phi$ mesons producing kaons (up to $41\%$ and $3.5\%$, respectively); (iii) the $k_{\mathrm T}$ dependences of the fragmentation functions, which can be different for different hadrons and can have an effect on the extracted amplitudes through the convolution of Eqs.~(\ref{eq:QPM-collins}) and (\ref{eq:convolution}).

\begin{figure}[!htb]
 \centering
{\includegraphics[height=0.296\textwidth,angle=0]{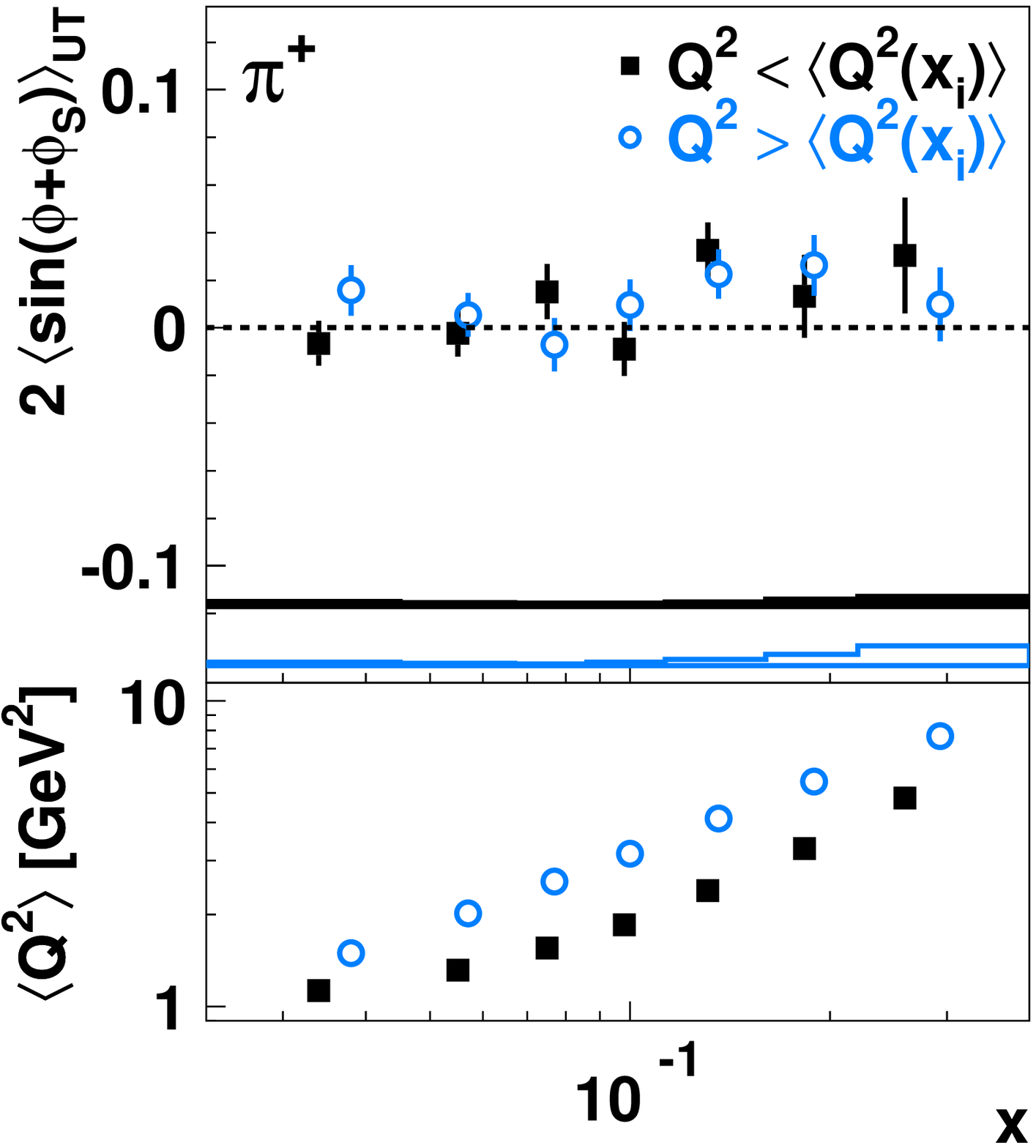}}
{\includegraphics[height=0.296\textwidth,angle=0]{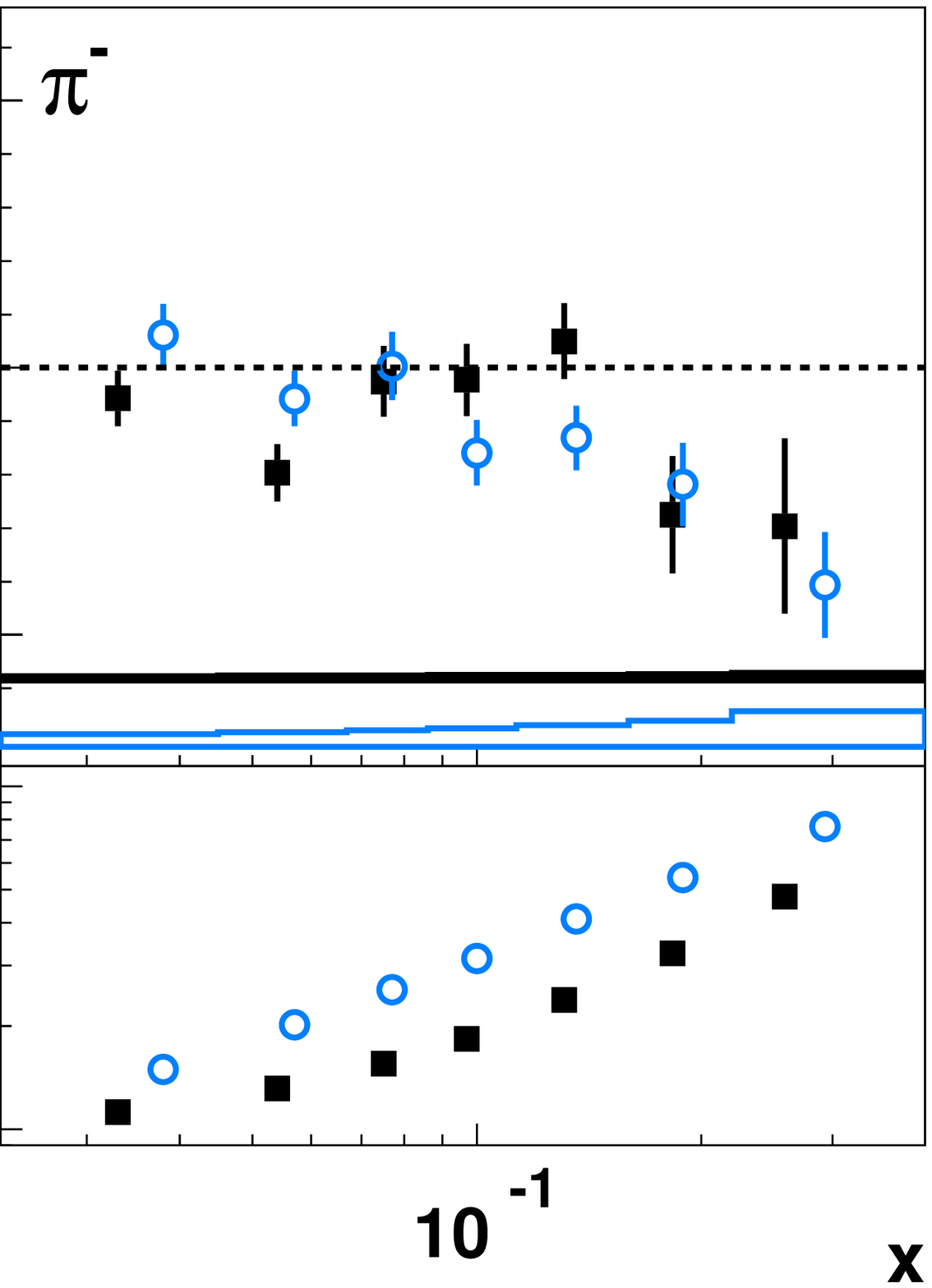}}
 \caption{Collins amplitudes for charged pions as functions of \(x\). The $Q^2$ range for each $i$-bin in $x$ was divided into the two regions above and below the average $Q^2$ of that bin (\(\langle Q^2(x_i)\rangle\)). The bottom panels show the $x$-dependence of the average \( Q^2\).}
\label{fig:Q2study}
\end{figure}

Up to this point, the discussion is based on Eq.~(\ref{eq:QPM-collins}) and is thus valid up to twist-3. It is therefore interesting to investigate the possible presence of twist-4 contributions. To this end, the \(Q^2\) dependence of the extracted amplitudes was studied in more detail. To minimize effects arising from the strong correlation between $x$ and $Q^2$ in the data, the events in each $x$ bin were divided into two sub-bins, with $Q^2$ below and above the mean value $\langle Q^2 (x_i) \rangle$ for the original bin (see Fig.~\ref{fig:Q2study}). However, due to the limited statistics it was not possible to significantly constrain the twist-4 contributions by fitting the data in Fig.~\ref{fig:Q2study} with various $Q^2$ dependences (including the appropriate $y$-dependent prefactor of Eq.~\ref{eq:QPM-collins}).


In summary, non-zero Collins amplitudes in semi-inclusive DIS were measured for charged pions and positively charged kaons. These amplitudes can be interpreted as due to the transverse polarization of quarks in the target, revealed by its influence on the fragmentation of the struck quark. They thus support the existence of non-zero transversity distribution functions in the proton and also the existence of non-zero Collins fragmentation functions. In particular, by comparing the Collins amplitudes of $\pi^+$ and  $\pi^-$, it appears that fragmentation that is disfavored in terms of quark flavor has an unexpected importance, and enters with a sign opposite to that of the favored one. In contrast to the expectation that the $\pi^+$ and the $K^+$ Collins amplitudes should have similar magnitudes, based on the common $u$-quark dominance, the amplitude for $K^+$ is found to be significantly larger than that for $\pi^+$. This could be an indication of, e.g, an unanticipated behavior of the Collins fragmentation functions possibly in conjunction with a non negligible role of the sea quarks in the nucleon. Collins amplitudes consistent with zero are measured for $\pi^0$ and $K^-$. These data should considerably improve the precision of transversity extractions from future global fits.


We gratefully acknowledge the \desy\ management for its support and the staff
at \desy\ and the collaborating institutions for their significant effort.
This work was supported by 
the Ministry of Economy and the Ministry of Education and Science of Armenia;
the FWO-Flanders and IWT, Belgium;
the Natural Sciences and Engineering Research Council of Canada;
the National Natural Science Foundation of China;
the Alexander von Humboldt Stiftung,
the German Bundesministerium f\"ur Bildung und Forschung (BMBF), and
the Deutsche Forschungsgemeinschaft (DFG);
the Italian Istituto Nazionale di Fisica Nucleare (INFN);
the MEXT, JSPS, and G-COE of Japan;
the Dutch Foundation for Fundamenteel Onderzoek der Materie (FOM);
the Russian Academy of Science and the Russian Federal Agency for 
Science and Innovations;
the U.K.~Engineering and Physical Sciences Research Council, 
the Science and Technology Facilities Council,
and the Scottish Universities Physics Alliance;
the U.S.~Department of Energy (DOE) and the National Science Foundation (NSF);
and the European Community Research Infrastructure Integrating Activity
under the FP7 "Study of strongly interacting matter (HadronPhysics2, Grant
Agreement number 227431)".


\bibliography{collins}
\bibliographystyle{elsart-num}

\end{document}